\documentclass[sigconf, natbib=false]{acmart}

\newcommand{\revision}[1]{{\color{blue}#1}}

\usepackage{mdframed}
\usepackage{xcolor}
\definecolor{light-gray}{gray}{0.95}

\usepackage{moreverb}
\usepackage{graphicx}
\usepackage{graphics}
\usepackage{subcaption}
\usepackage{textcomp}
\usepackage{colortbl}
\usepackage{multirow}
\usepackage{multicol}
\usepackage{url}
\usepackage{booktabs}
\usepackage{framed}
\usepackage{color}
\usepackage{xspace}
\usepackage{lstcustom}
\usepackage{algorithm}
\usepackage{algpseudocode}

\newcommand\BibTeX{{\rmfamily B\kern-.05em \textsc{i\kern-.025em b}\kern-.08em
T\kern-.1667em\lower.7ex\hbox{E}\kern-.125emX}}

\usepackage{lineno}

\usepackage{booktabs}
\usepackage{multirow}
\usepackage{graphicx}

\newcommand{\lstbg}[3][0pt]{{\fboxsep#1\colorbox{#2}{\strut #3}}}

\usepackage{enumitem}
\usepackage{parcolumns}
\usepackage{listings}
\lstdefinelanguage{diff}{
  basicstyle=\ttfamily\small,
  morecomment=[f][\lstbg{red!20}]-,
  morecomment=[f][\lstbg{green!20}]+,
  morecomment=[f][\textit]{@@},
  morecomment=[f][\textit]{---},
  morecomment=[f][\textit]{+++},
}

\newcommand{\gpt}{\textsc{ChatGPT4}}
\newcommand{\gemini}{\textsc{Gemini Advanced 1.5}}
\newcommand{\claude}{\textsc{Claude 3.5 Sonnet}}
\newcommand{\mistral}{\textsc{Le Chat Mistral}}

\usepackage{tikz}

\newcommand{\pie}[1]{%
\begin{tikzpicture}
 \draw (0,0) circle (1ex);\fill (1ex,0) arc (0:#1:1ex) -- (0,0) -- cycle;
\end{tikzpicture}%
}

\renewcommand{\revision}[1]{#1}

\AtBeginDocument{%
  \providecommand\BibTeX{{%
    Bib\TeX}}}

\setcopyright{acmcopyright}
\copyrightyear{2018}
\acmYear{2018}
\acmDOI{10.1145/1122445.1122456}

\acmConference[SBES'24]{Brazilian Symposium on Software Engineering}{September 30 -- October 04,
  2024}{Curitiba, PR}

\setcopyright{none} % This command was added for submissions to VEM
\renewcommand\footnotetextcopyrightpermission[1]{} % This command was added for submissions to VEM

\RequirePackage[
  datamodel=acmdatamodel,
  style=acmnumeric,
  ]{biblatex}

\begin{document}

%\title{Evaluating LLMs Ability to Detect Refactorings}
\title{Evaluating the Capability of LLMs in Identifying Compilation Errors in Configurable Systems}

 \author{Lucas Albuquerque}
 \orcid{0009-0003-6870-7782}
 \affiliation{%
   \normalsize \institution{Federal University of Campina Grande} \country{Brazil}
 }
 \email{lucas.albuquerque@ccc.ufcg.edu.br}

 \author{Rohit Gheyi}
 \orcid{0000-0002-5562-4449}
 \affiliation{%
   \normalsize \institution{Federal University of Campina Grande} \country{Brazil}
 }
 \email{rohit@dsc.ufcg.edu.br}
 
 \author{Márcio Ribeiro}
 \orcid{0000-0002-4293-4261}
 \affiliation{%
   \normalsize \institution{Federal University of Alagoas} \country{Brazil}
 }
 \email{marcio@ic.ufal.br}

\begin{abstract}

 %context
Compilation is an important process in developing configurable systems, such as Linux. 
% problem
However, identifying compilation errors in configurable systems is not straightforward because traditional compilers are not variability-aware. Previous approaches that detect some of these compilation errors often rely on advanced techniques that require significant effort from programmers. 
% solucao 
This study evaluates the efficacy of Large Language Models (LLMs), specifically \gpt{}, \mistral{} and \gemini{}, in identifying compilation errors in configurable systems.
% avaliação
Initially, we evaluate 50 small products in C++, Java, and C languages, followed by 30 small configurable systems in C, covering 17 different types of compilation errors. 
%The two LLMs were assessed based on their ability to correctly recognize and diagnose errors. 
% resultados
\gpt{} successfully identified most compilation errors in individual products and in configurable systems, while \mistral{} and \gemini{} detected some of them.
% conclusões
LLMs have shown potential in assisting developers in identifying compilation errors in configurable systems.%, specific challenges remain, particularly in configurable systems environments with high variability. %The study highlights the need for continuous refinement of LLMs to enhance their accuracy and utility in complex software development scenarios.

\end{abstract}

%    <concept>
%        <concept_id>10011007.10011006.10011073</concept_id>
%        <concept_desc>Software and its engineering~Software maintenance tools</concept_desc>
%        <concept_significance>500</concept_significance>
%        </concept>
%  </ccs2012>

%\ccsdesc[500]{Software and its engineering~Software maintenance tools}

\keywords{LLMs, Compilation Errors, Configurable Systems.}

\maketitle

\section{Introduction}
\label{sec:introduction}

\revision{
% compilacao
Compilation is an important process for creating functional and efficient programs. 
% sistemas configuráveis
This challenge is amplified in configurable systems, as seen with the Linux kernel, where variability and the combination of different modules and features can result in an exponential explosion of possible configurations. In such environments, finding bugs that occur only under specific configurations becomes a particularly costly and labor-intensive task. Developing configurable systems with dozens of macros is not easy~\cite{Medeiros15}, especially when annotations are not disciplined~\cite{Malaquias:2017,liebig-2010}, potentially affecting code quality~\cite{DBLP:conf/wcre/BaxterM01}. 
% problema
Traditional compilers can only check one configuration at a time. Variability-aware parsers are advanced parsing tools designed to handle software systems with multiple configurations and variability. Using the current variability-aware parsers~\cite{typechef,superc} is time consuming, require some effort to setup and do not detect all errors.

% contexto
%Compilation is an important process in software development, transforming source code written in high-level programming languages into machine code executable by a computer. This process is vital for creating functional and efficient programs. 
%However, errors often occur during compilation, and the error messages provided by compilers can be cryptic and uninformative, making debugging challenging for developers.
% contexto LLMs
Large Language Models (LLMs) have proven to be valuable tools in software generation and review, assisting with code writing and documentation~\cite{Goodfellow-et-al-2016, DBLP:conf/nips/VaswaniSPUJGKP17}. Some studies are investigating the extent to which LLMs can assist in testing activities~\cite{teste-llms-2024} and software engineering~\cite{se-llms-2023}. 
% problema em aberto
However, to the best of our knowledge, no study has yet explored the extent to which LLMs can aid in detecting variability-aware compilation errors.
}

In this paper, we evaluate the capability of Large Language Models (LLMs) in identifying compilation errors across diverse programming contexts. Our focus is to analyze the performance of three specific LLMs, \gpt{}, \mistral{} and \gemini{}, in identifying compilation errors. These LLMs were chosen because they are state-of-the-art models in the field, representing the latest advancements in large language model technology. Initially, we assessed the ability of these LLMs to identify issues in a set of 50 small programs across C++, Java, and C. Later, we expanded our analysis to include applying these models to 30 small configurable systems, ranging from 1 to 5 macros in C, with up to 33 LOC, examining 17 different types of compilation errors. 
% dados
All experimental data are available online~\cite{artefatos}.
% contribuições
In summary, our main contribution is the following:
\begin{itemize}
    \item Evaluate to what extent \gpt{}, \mistral{} and \gemini{} detect compilation errors in programs and configurable systems (Sections~\ref{sec:products}~and~\ref{sec:configurable}).
\end{itemize}
%This paper is structured as follows. Sections~\ref{sec:products}~and~\ref{sec:configurable} describes the methodology and results obtained from evaluating individual products and configurable systems, respectively. The subsequent sections cover related work and the study's conclusions, respectively.

\section{Evaluation: Products}
\label{sec:products}

First we assess LLM's performance in compiling single products.

\subsection{Methodology}

%Next we present the methodology used to evaluate the detection of compilation errors in individual products.

\subsubsection{GQM}

We structured our evaluation using the Goal-Question-Metric (GQM) approach~\cite{Basili1994}. The objective is to assess the effectiveness of LLMs, specifically \gpt{} and \mistral{}, in identifying compilation errors from the developers' perspective in the context of individual products.
We address the following research questions (RQs) to achieve the goal:
\begin{itemize}%
\item[\textbf{RQ$_{1}$}] To what extent can \gpt{} detect compilation errors in individual products? 
\item[\textbf{RQ$_{2}$}] To what extent can \mistral{} detect compilation errors in individual products?
\item[\textbf{RQ$_{3}$}] To what extent can \gemini{} detect compilation errors in individual products?
\end{itemize}
Each LLM's response will be compared to the language compiler to determine the number of correct and incorrect identifications.

\subsubsection{Planning}
\label{sec:products-planning}

The study's planning involves a structured methodology to assess the capabilities of the selected LLMs. The plan is as follows.
%\textbf{Product Selection.} 
% quantidade e quando
The study uses a sample of 50 products selected in April 2024 to ensure the results' relevance and timeliness. 
% local e tamanho
These products are divided between self-developed creations and code samples extracted from the Codeforces platform, distributed across C++, Java, and C, ranging from 7 to 70 lines of code (median: 25.06 LOC, mean: 24 LOC).
% explicacao Codeforces
The platform Codeforces was chosen for extracting code samples because it offers a wide variety of coding problems and solutions in multiple programming languages, ensuring diversity and real-world relevance. The code samples were chosen at random to mitigate selection bias and provide a realistic assessment of the LLMs' capabilities in handling typical programming errors.

% construcoes
The code snippets include (nested) loops, (nested) conditionals, functions, data structures (such as maps, arrays, and vectors), input and output operations, and mathematical calculations.
% erros de compiacao
Each product contains exactly one type of compilation error, which may include one to three instances of the same error. This selection aims to provide a comprehensive and representative analysis of LLM capabilities across different programming contexts.

%\textbf{Prompt Formulation.} 
The prompt used is ``Does the following \texttt{language} code compile? \texttt{code},'' where \texttt{language} specifies the programming language (C++, Java, or C) and \texttt{code} is the specific code snippet being analyzed. This prompt formulation was chosen for its simplicity, enabling a direct and focused interaction with the LLMs, specifically assessing their ability to determine whether the provided code compiles. \revision{We used default parameters.}
%\textbf{Compilation and Validation.} 
After receiving the LLMs' responses, each product is compiled using the appropriate compiler (GNU GCC 11 for C, GNU G++ 13 for C++, and Java 21 for Java). This step serves to validate the LLM responses against the compiler's verdict, which acts as a baseline for evaluation.

%\textbf{Qualitative Response Analysis.} 
The responses provided by the LLMs are analyzed based on five main criteria, where each response is classified as ``Yes,'' ``No,'' or ``Partially.'' ``Yes" indicates success, ``No'' denotes failure, and ``Partially'' ( $\pie{180}$ ) is used for detailed discussion in cases where success is not fully achieved but is considered a failure for final evaluation. The criteria are detailed as follows:
\begin{itemize}
    \item \textbf{Detect.} Determines if the LLM identified the presence of a compilation error.
    \item \textbf{Fix.} Evaluates if the LLM proposed an appropriate fix for the compilation error. The LLM should provide corrected code or directly and clearly describe a solution without changing the code's original purpose.
    \item \textbf{Explanation.} Assesses if the LLM satisfactorily explains the problem. Success is only considered if all sub-criteria are marked ``Yes,'' which includes:
    \begin{itemize}
        \item \textbf{Code Element.} Checks if the LLM pinpointed the specific code element causing the error.
        \item \textbf{Type of Error.} Determines if the LLM accurately classified the type of error.
        \item \textbf{Location.} Confirms if the LLM correctly indicated the error's location in the code. The LLM must specify in which function the error occurs, or, in the case of variables, identify the specific variable where the error happens.
    \end{itemize}
\end{itemize}
%\textbf{Models Evaluated.} 
In April 2024, we analyzed \gpt{} and \mistral{}. In May 2024, we also evaluated \gemini{}.

\subsection{Results}

The performance results of \gpt{},
 \mistral{}, \revision{and \gemini{}} are summarized in Table~\ref{tab:summary-products}.
%, highlighting significant differences in their ability to detect, correct, and explain compilation errors across various test scenarios.
\gpt{} exhibited a good performance in detecting and correcting errors, achieving 41 detections and 44 corrections out of a possible 50. In terms of explanation, this model was effective in 31 out of the 41 errors it detected.
\mistral{}, on the other hand, detected 28 errors and corrected 32 out of 50 products. \mistral{} explained 23 of the 28 errors it detected. It has a less consistent level compared to \gpt{}.
\revision{\gemini{} detected 27 errors, corrected 35 out of 50 products, and explained 21 of the 27 errors it detected. Among the three models, Gemini Advanced 1.5 was the least effective in error detection, although it showed a higher number of corrections than Le Chat Mistral.}

% \begin{table}[]
% \caption{Performance of LLMs in detecting, correcting, and explaining compilation errors in products.}
% \label{tab:llms-performance}
%
%\begin{tabular}{|r|r|r|r|}
% \hline
% \rowcolor[HTML]{333333} 
% \multicolumn{1}{|c|}{\cellcolor[HTML]{333333}{\color[HTML]{FFFFFF} \textbf{LLM}}} & \multicolumn{1}{c|}{\cellcolor[HTML]{333333}{\color[HTML]{FFFFFF} \textbf{Detect}}} & \multicolumn{1}{c|}{\cellcolor[HTML]{333333}{\color[HTML]{FFFFFF} \textbf{Fix}}} & \multicolumn{1}{c|}{\cellcolor[HTML]{333333}{\color[HTML]{FFFFFF} \textbf{Explanation}}} \\ \hline
% GPT4                                                                              & 41/50                                                                               & 44/50                                                                            & 31/41                                                                                    \\ \hline
% Mistral                                                                           & 28/50                                                                               & 32/50                                                                            & 23/28                                                                                    \\ \hline
% \end{tabular}
% \end{table}

\begin{table*}[]
\caption{Evaluation results of identifying compilation errors in products.}
\label{tab:summary-products}
\resizebox{\textwidth}{!}{%
\begin{tabular}{rcrcl|ccccc|ccccc|ccccc|}
\cline{6-20}
\multicolumn{1}{l}{}                                                             & \multicolumn{1}{l}{}                                                                  & \multicolumn{1}{l}{}                                                             & \multicolumn{1}{l}{}                                                              &                                                      & \multicolumn{5}{c|}{\cellcolor[HTML]{000000}{\color[HTML]{FFFFFF} \textbf{\gpt{}}}}                                                                                                                                                                                                                                                                                                     & \multicolumn{5}{c|}{\cellcolor[HTML]{000000}{\color[HTML]{FFFFFF} \textbf{\mistral{}}}}                                                                                                                                                                                                                                                                                                 & \multicolumn{5}{c|}{\cellcolor[HTML]{000000}{\color[HTML]{FFFFFF} \textbf{\gemini{}}}}                                                                                                                                                                                                                                                                                                  \\
\hline
\end{tabular}%

}
\end{table*}

%In syntactic errors like ``Missing semicolon'' and ``Mismatching quotes,'' both models showed high success rates, as detailed in Table~\ref{tab:summary-products}. However, in more complex scenarios such as ``Mismatching brackets'' and ``Type mismatch,'' \gpt{} outperformed \mistral{}. \mistral{} failed completely, such as in detecting ``Mismatching brackets''.

\subsection{Discussion}

%In this section, we analyze the results achieved. 

\subsubsection{Compilation Error Detection}

% os tres juntos
In six examples, none of the three LLMs can detect the compilation errors. However, in four of these cases, at least one LLM suggests a code improvement that resolves the compilation error.
% tipos
When analyzing the specific types of compilation errors detected by the LLMs, we observe varying performance across different error categories. Notably, syntax errors like ``Missing semicolon'' and ``Mismatching parentheses'' exhibited high detection rates by all three LLMs, with \gpt{} identifying all 8 cases of ``Missing semicolon'' and \mistral{} detecting 7 of them. \revision{Detection of ``Mismatching brackets'' was particularly strong in \gpt{}, detecting 6 out of 6 cases, while \mistral{} and \gemini{} detected none.}
In contrast, semantic errors like ``Variable not declared'' and ``Type mismatch'' proved more challenging, with both LLMs showing moderate results. \gpt{}, \mistral{}, and \gemini{} each detected 3 out of 4 ``Variable not declared'' cases.

%This variability in the LLMs' performance suggests that while both are useful tools for identifying compilation errors, there are significant differences in how each model processes and responds to different types of errors. This could be due to differences in their underlying architectures, training sets, or how specific programming language features are interpreted and analyzed by each model.

One specific error analyzed during the evaluation was ``variable out of scope,'' where \gpt{} correctly identified 7 out of 10 instances. LLMs can identify variables used outside their permissible scope. However, the two instances where errors were not detected involved a common scenario in C++ programming: the declaration of a variable within a for loop header and attempting to access this variable immediately after the loop ends (Id 14 from Table~\ref{tab:summary-products}). 
%This type of error is illustrated in Figure~\ref{fig:gpt-does-not-detect}, demonstrating a specific area where \gpt{} error detection needs improvement.
% conclusao
LLMs have 70\%{} success rate in detecting out-of-scope variables. However, the difficulty in identifying errors involving scopes limited to specific blocks, such as those introduced by loops, suggests an opportunity for improvement.

% \begin{figure*}[ht]
% \begin{center}
% \leavevmode
% \scalebox{0.42}{
% \includegraphics{images/gpt-does-not-detect.png}}
% \caption{
% \gpt{} does not detect the compilation error but ends up fixing it.}
% \label{fig:gpt-does-not-detect}
% \end{center}
% \end{figure*}

\subsubsection{Compilation Error Fixing}

In some cases where LLMs did not explicitly detect an error, they still suggested changes to the code. Interestingly, these proposed changes, although not initially aimed at fixing a specific identified issue, ended up resolving the problem. This led to a number of effective corrections exceeding the detected compilation errors.
% exemplo
For instance, \gpt{} did not initially identify an error in the for element (the missing closing parenthesis) in Id 3: \texttt{for (int i = 0; i < (int)a.size(); i++}. But, it suggests to use a range-based \texttt{for} loop to simplify the code: \texttt{for (int num : a)}. So, it fixed the compilation error.
% mistral
On the other hand, \mistral{} \revision{and \gemini{}} not only detect the compilation error:
\begin{mdframed}[backgroundcolor=light-gray, linecolor=black, linewidth=0.5pt]
\textit{``... There is a missing right parenthesis `)' in the for loop declaration ...''}
\end{mdframed}
but also provide a fix.

%An example is illustrated in Figure~\ref{fig:fix-without-detect}. 
%\gpt{} did not initially identify an error in the for element (the missing closing parenthesis), but suggested changing to a range-based loop, which not only simplifies the code but also eliminates the need for type casting, implicitly resolving the compilation problem (Id 3 from Table~\ref{tab:summary-products}). This suggestion not only improves the code's clarity and maintainability but also avoids potential errors arising from the improper use of data types.

% \begin{figure}[ht]
% \begin{center}
% \leavevmode
% \scalebox{0.27}{
% \includegraphics{images/fix-without-detecting.png}}
% \caption{
% \gpt{} does not detect the compilation error but ends up fixing it.}
% \label{fig:fix-without-detect}
% \end{center}
% \end{figure}

\subsubsection{Explanation}
%The explanation of compilation errors provided by LLMs, such as \gpt{} and \mistral{}, reveals significant aspects of the interaction between the model's understanding and the presentation of accurate information. 
While hallucinations, or the generation of incorrect and fictitious information by LLMs, are a known issue~\cite{alucinacao}, the results of this evaluation show that the models often provide coherent and useful explanations, even in cases where the initial detection may seem uncertain.
% resultados
In the results presented, we observed that in some instances, LLMs initially indicate no compilation errors, but as the response develops, they recognize the presence of issues, adjusting their initial conclusions. Although this change in stance might seem inconsistent, it rarely compromises the quality of the explanations provided. The final responses, which include corrections to the model's initial assessment, typically offer detailed explanations of the nature and context of the detected error.
% exemplo
% For example, \gpt{} successfully provided satisfactory explanations for 31 out of the 41 errors it detected, and \mistral{} adequately explained 23 out of the 28 errors it identified. These numbers reflect a relatively high success rate in correctly explaining issues.
% conclusão
According to our ``Explanation'' evaluation metrics, the results were considered satisfactory, with all three LLMs successfully meeting the explanation criteria more than 75\% of the time. 
%This is crucial in real-world software development environments, where accurate information is vital.

\subsection{Threats to Validity}

\revision{A factor that can compromise result validity is selection bias in code samples, as choosing examples that don't adequately represent the diversity of real-world errors could skew the evaluation of LLMs. We created some examples based on compilation errors found in real systems. The modifications made and the new examples created help to minimize the risk of data leakage when using LLMs~\cite{threats-llms-icse-nier-2024}.}
%Furthermore, while the compilers used as the baseline are generally reliable, they are not free from bugs. To mitigate this risk, each code sample was manually verified by the author to ensure the accuracy of the compilation data. 
%This approach seeks to minimize threats to the study's validity and guarantee the correct interpretation of results.
\section{Evaluation: Configurable Systems}
\label{sec:configurable}

%In this section, we present the evaluation for detecting compilation errors in configurable systems.
Next, we evaluate configurable systems.

\subsection{Methodology}

%Next we outline the methodology used to evaluate the detection of compilation errors in configurable systems.

\subsubsection{GQM}

The objective is to assess the effectiveness of LLMs, specifically \gpt{} and \mistral{}, in identifying compilation errors from the developers' perspective in the context of configurable systems.
We address the following RQs:

\begin{itemize}%
\item[\textbf{RQ$_{1}$}] To what extent can \gpt{} detect compilation errors in configurable systems? 
\item[\textbf{RQ$_{2}$}] To what extent can \mistral{} detect compilation errors in configurable systems? 
\item[\textbf{RQ$_{3}$}] To what extent can \gemini{} detect compilation errors in configurable systems?
\end{itemize}
Each LLM's response will be compared to the language compiler for each product within the configurable system to accurately determine the number of correct and incorrect identifications.

\subsubsection{Planning}

The study's planning involves a structured methodology to assess the capabilities of the selected LLMs. 
%\textbf{Product Selection.} 
% tamanho
We included 30 configurable systems, ranging from 4 to 33 LOC (median: 16.8 LOC, mean: 16 LOC). 
% macros
Each configurable system contains 1 to 5 macros and contains one or two types of compilation errors. 
% construcoes codigo
The code snippets include loops, conditionals, functions, data structures (such as maps, arrays, and vectors), input and output operations, and mathematical calculations. Additionally, they feature nested ifdefs, ifdefs with simple boolean expressions, both disciplined and undisciplined ifdefs~\cite{liebig-2010}, ifdefs inside functions, and ifdefs within function declarations.
% local
We created 14 configurable systems. Additionally, there are six systems that are based on Braz et al.'s studies that identified compilation errors in configurable systems~\cite{Braz:2016, Braz:2018}. The remaining systems are adapted from Abal et al.'s research on variability bugs in the Linux kernel, providing simplified versions of the original code~\cite{abal-2014,Abal18}.

%\textbf{Prompt Formulation.} 
We used the prompt ``Does the following C code compile? \texttt{code},'' where \texttt{code} represents the code snippet. This prompt was chosen for simplicity, focusing on direct interaction with the LLMs to evaluate their ability to comprehend and process conditional compilation. English was used because LLMs are trained on a significantly larger volume of data in this language.
%\textbf{Compilation and Validation.} 
Each configurable system is compiled using the GNU GCC 11 compiler for C. During this process, we manually analyzed how many unique products could be generated by activating different features. Each unique product was manually compiled to verify how many configurations contained compilation errors. The 30 configurable systems generated a total of 103 unique products, of which 40 contained compilation errors.

%\textbf{Qualitative Response Analysis.} 
The analysis of the LLM responses follows the same structure used in the evaluation of individual products (Section~\ref{sec:products-planning}), with specific adaptations for the configurable systems context:

\begin{itemize}
    \item \textbf{Detect.} It refers to the number of distinct products with compilation errors that the LLMs successfully identified.
    
    \item \textbf{Fix.} To classify a correction as ``Yes,'' the LLM's proposed solution must be general and applicable to all products, without relying on specific adjustments like directly defining macros in the code that only guarantee compilation in that particular configuration. This criterion seeks to evaluate the model's ability to propose sustainable and generalizable fixes that maintain product functionality without specific manual interventions.
    
    \item \textbf{Explanation.} The explanation evaluation follows the same approach as the individual product analysis, considering whether the LLM can satisfactorily clarify the detected problem. This includes correctly identifying the code element causing the compilation error, the nature of the error, and the specific location of the issue within the code of a product.
\end{itemize}
%\textbf{Models Evaluated.}
\revision{
In April 2024, we analyzed \gpt{} and \mistral{}. In May 2024, we evaluated \gemini{}. We utilized the default parameters.
}

\subsection{Results}

% resumo
The results of the configurable systems evaluation using \gpt{}, \revision{\gemini{}} and \mistral{} are presented in Table~\ref{tab:summary-configurable-systems}.
% detalhes
\gpt{} detects all compilation errors (CE) in 28 of the 30 tested configurable systems. This model also identified errors in 38 of the 40 individual configurations derived from these lines, missing only ``Type Mismatch'' and ``Variable not declared.'' In Id 19, \gpt{} detected errors in only one of the two erroneous configurations. In terms of fixes, \gpt{} proposed effective fixes for 12 of the 30 configurable systems. Regarding explanations, the model was able to provide adequate explanations for 26 of the 29 configurable systems with compilation errors.

\begin{table*}[]
\caption{Evaluation results of identifying compilation errors in configurable systems.}
\label{tab:summary-configurable-systems}
\resizebox{\textwidth}{!}{%

\begin{tabular}{lllllll|r|ccccc|ccccc|ccccc|}
\cline{8-23}
\rowcolor[HTML]{000000} 
\cellcolor[HTML]{FFFFFF}{\color[HTML]{000000} }                         & \cellcolor[HTML]{FFFFFF}{\color[HTML]{000000} }                           & \cellcolor[HTML]{FFFFFF}{\color[HTML]{000000} }                            & \cellcolor[HTML]{FFFFFF}{\color[HTML]{000000} }                            & \cellcolor[HTML]{FFFFFF}{\color[HTML]{000000} }                         & \cellcolor[HTML]{FFFFFF}{\color[HTML]{000000} }                              & \cellcolor[HTML]{FFFFFF}{\color[HTML]{000000} }              & \multicolumn{1}{l|}{\cellcolor[HTML]{000000}{\color[HTML]{FFFFFF} }} & \multicolumn{5}{c|}{\cellcolor[HTML]{000000}{\color[HTML]{FFFFFF} \textbf{\gpt{}}}}                                                                                                                                                                                                                                                                                                                                       & \multicolumn{5}{c|}{\cellcolor[HTML]{000000}{\color[HTML]{FFFFFF} \textbf{\mistral{}}}}                                                                                                                                                                                                                                                                                                                                   & \multicolumn{5}{c|}{\cellcolor[HTML]{000000}{\color[HTML]{FFFFFF} \textbf{\gemini{}}}}                                                                                                                                                                                                                                                                                                                                    \\
\hline
\end{tabular}%

}
\end{table*}

\mistral{}, meanwhile, identified all compilation errors in 24 of the 30 configurable systems, and 31 of the 40 individual configurations. The model managed to propose fixes for 9 of the 30 configurable systems and provided adequate explanations for 18 of the 26 detected lines. In Ids 2 and 21, \mistral{} detected compilation errors in some configurations but not all.
%\mistral{} cannot detect the following errors: Type not declared (2), Function redefinition (1), Function signature mismatch (4), Variable not declared (1), and Invalid return type (1).
%A summary of the compilation errors missed by \mistral{} can be found in Table~\ref{tab:llms-performance-config-not-detected-mistral}. 
These results indicate that although it is effective at detecting some compilation errors, \mistral{} faces more challenges in proposing effective fixes and providing detailed explanations.

\revision{
\gemini{}, on the other hand, can detect all compilation errors in 16 configurable systems. However, it incorrectly states that 10 configurable systems do not have compilation errors. In four configurable systems, it detects some of the compilation errors. In our study, the \textsc{Gemini} performance is worse than \gpt{} and \mistral{}.
}

% \begin{table}[]
% \caption{Compilation errors undetected by \mistral{}.}
% \label{tab:llms-performance-config-not-detected-mistral}

% \begin{tabular}{|c|c|}
% \hline
% \rowcolor[HTML]{333333} 
% {\color[HTML]{FFFFFF} \textbf{Compilation Error}} & {\color[HTML]{FFFFFF} \textbf{\#Not Detected}} \\ \hline
% Type not declared                     & 2                                              \\ \hline
% Function redefinition                 & 1                                              \\ \hline
% Function signature mismatch           & 4                                              \\ \hline
% Variable not declared                 & 1                                              \\ \hline
% Invalid return type                   & 1                                              \\ \hline
% \end{tabular}

% \end{table}

\subsection{Discussion}

%In this section, we analyze the results achieved.

\subsubsection{Compilation Error Detection}

Most undetected errors by LLMs are semantic: both undetected errors by \gpt{} and eight out of nine undetected errors by \mistral{} are semantic, pointing to a potential area for enhancement.
% exemplo
We present a configurable system (Id 10 from Table~\ref{tab:summary-configurable-systems}) in Listing~\ref{lst:example-mistral}. In this example, the \texttt{norm} function adapts its calculations depending on whether the macros \texttt{A} and \texttt{B} are defined. When macro \texttt{A} is not defined, the \texttt{norm} function is configured to accept only two parameters, conflicting with the call made in the main function, where \texttt{norm} is invoked with three arguments.
\begin{lstlisting}[basicstyle=\footnotesize,language=C, label=lst:example-mistral, caption={\mistral{} \revision{ and \gemini{}} do not detect a compilation error in Id 10.}]
#include <stdio.h>
struct point { int x, y;};
int norm(
        int x,
#ifdef A
        int y,
#endif
        int z
        ) {
    int w = x * z;
#ifdef B
    w += y;
#endif
    return w;
}
int main() {
    int x = norm(1, 2, 3);
    printf("%d\n", x);
    return 0;
}
\end{lstlisting}

\revision{
This inconsistency should lead to a compilation error, but both \mistral{} and \gemini{}, while analyzing possible products, identifies this only as a warning rather than a compilation error, failing to recognize the problem:
}
\begin{mdframed}[backgroundcolor=light-gray, linecolor=black, linewidth=0.5pt]
\textit{``... Neither \texttt{A} nor \texttt{B} defined: The code compiles without errors. The \texttt{norm} function has arguments \texttt{x}  and \texttt{z}, and it only calculates \texttt{w = x * z}. The \texttt{main} function calls \texttt{norm(1, 2, 3)}, passing the extra argument \texttt{2}, but it's ignored by the function ...''}
\end{mdframed}  % esse trecho é da resposta do gemini
\revision{
\gemini{} does point out another problem (Listing~\ref{lst:example-mistral}):
\begin{mdframed}[backgroundcolor=light-gray, linecolor=black, linewidth=0.5pt]
\textit{``... Only \texttt{B} defined: The code won't compile. There will be an error because \texttt{y} is used in the \texttt{norm} function but not declared as an argument ...''}
\end{mdframed}
% conclusão
This seems to indicate a lack of understanding of the C language by both \mistral{} and \gemini{}. On the other hand, \gpt{} detects both problems.

Furthermore, Listing~\ref{lst:example-gemini} shows an example containing a missing semicolon (Id 7).
\gpt{} and \mistral{} can detect and explain how to fix it. \textsc{Gemini} does not detect the compilation error:
}
\begin{mdframed}[backgroundcolor=light-gray, linecolor=black, linewidth=0.5pt]
\textit{``... if you were to define \texttt{A}  then the subtraction would be performed ...''}
\end{mdframed}
\begin{lstlisting}[basicstyle=\footnotesize,language=C, label=lst:example-gemini, caption={\gemini{} does not detect a compilation error in Id 7.}]
#include <stdio.h>
int main() {
    int a = 3, b = 7;
    int x = a * b;
#ifdef A
    x -= 1
#endif
    printf("%d\n", x);
}
\end{lstlisting}

\mistral{} can detect the compilation errors not detected by \gpt{} in two configurable systems containing nested ifdefs (Ids 19 and 27). There are other examples with nested ifdefs (Ids 20 and 21).
% Id 27
\gpt{} cannot detect a type mismatch in Id 27 when one macro is enabled and another one is disabled. 
% gpt4o
However, the latest \gpt{}-o detects and explains in details the compilation error.
%mistral
\mistral{} correctly explains why the compilation error happens.
% Id 19
In Id 19, we have 3 macros, and \gpt{} yields the following message:
\begin{mdframed}[backgroundcolor=light-gray, linecolor=black, linewidth=0.5pt]
\textit{``... If \texttt{B} is defined but \texttt{C} is not, the variable \texttt{x} is defined and updated in the loop, but \texttt{y} remains undeclared and unused ...''}
\end{mdframed}
It correctly indicates that \texttt{y} remains undeclared but \gpt{} cannot identify that \texttt{y} is used in \texttt{x += sum + y;}.
% gpt-4o
\gpt{}-o \revision{and \gemini{} also do not detect it}.
% mistral
On the other hand, \mistral{} correctly identifies the compilation error:
\begin{mdframed}[backgroundcolor=light-gray, linecolor=black, linewidth=0.5pt]
\textit{``... the variable \texttt{y} will not be declared, but there is an attempt to use \texttt{y} in the expression \texttt{x += sum + y;}, which will result in a compilation error ...''}
\end{mdframed}
%Based on the preliminary results, we observe that the LLMs, particularly \gpt{} seems to identify logical patterns within product lines instead of attempting brute-force approaches to validate the well-formedness of every configuration. An example of this capability is illustrated in Figure 5, where the function \texttt{f} depends on the definitions of macros \texttt{A}, \texttt{B}, and \texttt{C}. \gpt{} demonstrated understanding that if macro \texttt{A} is not defined, the \texttt{f} function and any code within its block will not be compiled, making further checks for \texttt{B} and \texttt{C} irrelevant for this compilation.

%This ability to recognize and prioritize macro dependencies avoids redundant analysis of configurations that do not affect the final compilation outcome, optimizing the error detection process. This reasoning not only improves analysis efficiency but also highlights the sophistication of the model in understanding and applying preprocessing logic, which is crucial for managing the complexity of software product lines.

\subsubsection{Compilation Error Fixing}
During the evaluation of configurable systems, LLMs suggested fewer fixes compared to individual products (Section~\ref{sec:products}). 
%An important observation is that while the LLMs detect and explain errors, their ability to suggest effective corrections was less consistent. 
% aplicacao atual
Despite this decline, the results are still promising, indicating that even in more complex scenarios, LLMs can identify and suggest valid interventions, albeit less frequently.
% conclusão
This suggests that although the models have a reasonable understanding of errors, they could benefit from more specific guidance when proposing solutions. 
% future work: nosso
This leads us to consider improvements for future work, where the prompt used to interact with the LLMs could be adjusted to explicitly request a correction. Changing the prompt to explicitly request a solution could help guide the models not only to identify the issue but also to focus more directly on generating an applicable fix.

\subsubsection{Explanation}
\gpt{} demonstrated a good performance in explaining errors, providing consistent and clear details that aid in understanding the issues detected. This model was effective in clarifying the contexts of compilation errors and their implications, especially for complex cases like ``Function not defined'' and ``Type not declared,'' where the explanations were detailed and informative. Additionally, \gpt{} responses were longer and more comprehensive than those of \mistral{}.

\revision{
Both \mistral{} and \gemini{} exhibited limitations in crafting detailed explanations. The model were less consistent, especially in cases requiring a deeper understanding of the interactions between multiple macros and their impact on the code's logic. The discrepancy between compilation error detection and the quality of explanations was more pronounced, indicating significant room for improvement in the accuracy and depth of responses.
All three LLMs struggled to explain errors involving configurable systems that generate 4$^{+}$ products, often failing to provide explanations that fully captured the nature and cause of the issues. This challenge suggests that, although useful, the models still require refinement to effectively handle the complexity of configurable systems.
}

\subsection{Threats to Validity}

% exemplos
Selection bias in the code samples is a significant concern since examples that don't adequately capture the diversity of errors found in real-world development environments can lead to a skewed evaluation of the LLMs' capabilities. 
% mitigar
We created some examples based on compilation errors found in real configurable systems.
% data leakage
The modifications made and the new examples created help to minimize the risk of data leakage when using LLMs~\cite{threats-llms-icse-nier-2024}.
% compiladores
Additionally, while the compilers used as a baseline are generally reliable, the possibility of them containing bugs cannot be completely ruled out. 
% mitigar
We manually analyze the compiler results.

Beyond these aspects, a specific limitation of this study was the relatively small size of configurable systems assessed, with the largest containing only 33 LOC. Many configurable systems are simplified versions of more complex codebases, potentially making it easier for the LLMs to detect and correct errors. This simplification might not fully reflect the challenges encountered in more extensive and intricate software scenarios, potentially inflating the models' perceived effectiveness.

\section{Related Work}
\label{sec:related}
% 1 parágrafo por trabalho
% o que o trabalho relacionado fez?
% o que nós fazemos a mais?
% o que eles fazem a mais?

Some variability-aware tools have been previously proposed, such as TypeChef~\cite{typechef} and SuperC~\cite{superc}, for detecting certain syntax and type errors in configurable systems written in C. These tools use advanced techniques to implement non-trivial static analyses to identify compilation errors in real-world configurable systems. Users must configure these tools before use. 
% relacionamento
Our work assesses how well LLMs can perform variation-aware analysis, requiring minimal effort from the user.
%, who only needs to provide a prompt and add the code for the configurable system.

Abal et al.~\cite{Abal18,abal-2014} identified a number of bugs in configurable C systems and studied their characteristics. Some of these bugs are related to compilation errors and were included in our work. 
% relacionamento
For future work, we aim to explore how well LLMs can identify other issues, such as vulnerabilities in addition to compilation errors, using this set of cataloged examples~\cite{Abal18,DBLP:conf/sigsoft/MordahlOKWG19}. For instance, we could evaluate a set of vulnerabilities in configurable systems identified by previous approaches~\cite{Muniz:2018,flavio-sbes-2020}.

Medeiros et al.~\cite{flavio-gpce-2013} proposed a technique to identify a set of syntax errors in configurable C systems. Later, Medeiros et al.~\cite{flavio-gpce2015} proposed a method to detect undeclared variable usage. Both techniques could identify real-world compilation errors in configurable C systems. 
% relacionamento
In our evaluation, some examples are associated with bugs found in these earlier studies.

Braz et al.~\cite{Braz:2016, Braz:2018} proposed a technique to detect compilation errors in configurable C systems by analyzing the impact of changes. They suggested a non-trivial static analysis to identify new compilation errors introduced by changes. This technique successfully identified multiple compilation errors in real systems. 
% relacionamento
Our work adopts a simpler approach by using LLMs to detect compilation errors in configurable systems. For future work, we plan to evaluate not only real systems but also LLMs with larger context windows like \textsc{Gemini} to handle larger examples.

\section{Conclusion}
\label{sec:conclusion}

In this paper, we evaluate the extent to which LLMs such as \gpt{} and \mistral{} are capable of identifying compilation errors in configurable systems. \gpt{} successfully identified 41 out of 50 possible errors in products and 28 out of errors in 30 \revision{configurable systems}, demonstrating high effectiveness in detecting compilation errors. On the other hand, \mistral{} identified 28 out of 50 errors in products and 24 out of 30 errors in configurable systems in small examples. \gemini{} identified errors in 16 out of 30 configurable systems.
LLMs have shown potential in assisting developers in identifying compilation errors in configurable systems.
\revision{Some of them are not detected by variability-aware parsers~\cite{typechef}. The \gpt{}'s explanations help developers to understand and fix them.}

\revision{\textbf{Future Work.}} We plan to evaluate real systems. Additionally, we intend to consider other LLMs such as \revision{\claude{}, GitHub Copilot,} Llama 3, among others. 
%We also plan to perform fine-tuning on models like Llama 3 to see to what extent we can achieve improved performance in error detection in configurable systems. Another important aspect is to conduct metamorphic testing~\cite{metamorphic-testing} to mitigate potential threats to validity. 
We also aim to evaluate other prompts~\cite{prompts,prompt-techniques}, as well as assess how well LLMs can detect and correct compilation errors more deeply, especially in real configurable systems. \revision{We aim to investigate the time required for processing by LLMs and the extent to which they propose fixes.} We may face challenges similar to those encountered previously by other techniques analyzing highly configurable systems~\cite{Medeiros16}. We will consider the use of sampling algorithms in these scenarios in the context of LLMs.

\section*{Acknowledgments}
\revision{We would like to thank the anonymous reviewers for their insightful suggestions. 
This work was partially supported by CNPq and FAPEAL grants.}

%\bibliographystyle{ACM-Reference-Format}
%\bibliography{ref}

\printbibliography

\end{document}